\documentclass[journal,a4paper]{IEEEtran}
\usepackage[T1]{fontenc}
\usepackage{layouts}
\usepackage{cite}
\usepackage[pdftex,colorlinks=true,bookmarks=true,citecolor=blue,urlcolor=blue]{hyperref} 

\ifCLASSINFOpdf
  \usepackage[pdftex]{graphicx}
   \graphicspath{{./figs/}}
  \DeclareGraphicsExtensions{.pdf,.jpeg,.png}
\else
\fi

\usepackage{amsmath}
\usepackage{amsthm}
\usepackage[cmintegrals]{newtxmath}
\usepackage{bm}
\usepackage{algorithmic}
\usepackage{array}
\usepackage{mathrsfs}  
\usepackage{amsbsy}
\usepackage{amssymb}
\usepackage{amsthm}
\usepackage{bm}
\graphicspath{{./figs_nconst/}}

\newcommand{\field}[1]{\mathbb{#1}}

\newcommand{\sdots}{\reflectbox{$\ddots$}}

\DeclareMathOperator{\Wrons}{\mathscr{W}}

\newcommand{\tp}{\intercal}

\newcommand{\ovl}[1]{\overline{#1}}

\newcommand{\bigO}[1]{\mathop{\mathscr{O}}\left(#1\right)}

\newcommand{\floor}[1]{\left\lfloor #1 \right\rfloor}

\let\Re\relax
\DeclareMathOperator{\Re}{Re}
\let\Im\relax
\DeclareMathOperator{\Im}{Im}
\newcommand{\vv}[1]{\boldsymbol{#1}}
\newcommand{\vs}[1]{\boldsymbol{#1}}

\newcommand{\OP}[1]{\mathscr{#1}}

\newtheorem{rem}{Remark}[section]

\usepackage{enumitem}

\begin{document}
\title{A Chebyshev Spectral Method for Nonlinear Fourier 
Transform: Norming Constants}
\author{Vishal Vaibhav
\thanks{Email:~\tt{vishal.vaibhav@gmail.com}}
}

\maketitle
\begin{abstract}
In this paper, we present a Chebyshev based spectral method for the 
computation of the Jost solutions corresponding to complex values of the spectral 
parameter in the Zakharov--Shabat scattering problem. The discrete framework is 
then used to devise a new algorithm based on a minimum total 
variation (MTV) principle for the computation of the 
norming constants which comprise the discrete part of the nonlinear Fourier 
spectrum. The method relies on the MTV principle to find the points where 
the expressions for norming constants are numerically well-conditioned.
\end{abstract}

\IEEEpeerreviewmaketitle
\section{Introduction}
This paper considers the Zakharov and Shabat (ZS)~\cite{ZS1972} scattering 
problem which forms the basis for defining a nonlinear generalization of 
the conventional Fourier transform dubbed as the 
\emph{nonlinear Fourier transform} (NFT). In an 
optical fiber communication system the nonlinear Fourier (NF) spectrum 
offers a novel way of encoding information in optical pulses where the nonlinear effects 
are adequately taken into account as opposed to being treated as a source of 
signal distortion~\cite{Yousefi2014compact,TPLWFK2017}. One of the challenges 
that has emerged in realizing these ideas is the development of a fast and 
well-conditioned NFT algorithm that can offer spectral accuracy at low 
complexity. Such an algorithm would prove extremely useful for system design 
and benchmarking. Currently, there are primarily two successful 
approaches proposed in the literature for computing the continuous 
NF spectrum which are capable of achieving algebraic orders convergence 
at quasilinear complexity: (a) the integrating factor (IF) based exponential 
integrators~\cite{VW2017OFC,V2017INFT1,V2018BL,V2018LPT,V2019LPT} (b) exponential time 
differencing (ETD) method based exponential 
integrators~\cite{V2019CNFT}. Note that while the IF schemes uses fast 
polynomial arithmetic in the monomial basis, the ETD schemes 
use fast polynomial arithmetic in the Chebyshev basis. For the inverse 
transform, a sampling series based approach for computing the 
``radiative'' part has been proposed in~\cite{V2019BL1} which achieves spectral 
accuracy at quasilinear complexity per sample of the signal. In this paper, 
we extend the recently proposed spectral method~\cite{V2019SNFT} for the 
computation continuous spectrum to compute the norming constants. It is 
well-known that the determination of the point where the 
expression which defines the norming constant is numerically well-conditioned 
is non-trivial problem. In the previous works~\cite{V2017INFT1,V2018BL,V2018LPT} 
this point was taken to be origin, however, this choice can be shown to fail for 
carefully constructed examples. In order to remedy this problem, we propose 
a \emph{minimum total variation} (MTV) principle to determine a set of points 
where the expression for the norming constants are 
well-conditioned. Note that total variation of the 
quantities in question are identically zero at the continuous level; therefore, 
it makes sense to seek the minima of TV for a sliding window of fixed 
size which traverses the sampling grid. The size of the sliding window can 
be adaptively reduced which adds to the effectiveness of the algorithm.

We begin our discussion with a brief review of the scattering theory closely
following the formalism presented in~\cite{AKNS1974}. The nonlinear Fourier
transform of any signal is defined via the Zakharov-Shabat (ZS)~\cite{ZS1972} scattering 
problem which can be stated as follows: For $\zeta\in\field{R}$ and $\vv{v}=(v_1,v_2)^{\tp}$,
\begin{equation}\label{eq:ZS-prob}
\vv{v}_t = -i\zeta\sigma_3\vv{v}+U(t)\vv{v},
\end{equation}
where $\sigma_3$ is one of the Pauli matrices defined in the beginning of this
article. The potential 
$U(t)$ is defined by $U_{11}=U_{22}=0,\,U_{12}=q(t)$ and $U_{21}=-q^*(t)$. Here, 
$\zeta\in\field{R}$ is known as the \emph{spectral parameter}
and $q(t)$ is the complex-valued signal. The solution of the scattering 
problem~\eqref{eq:ZS-prob}, henceforth referred
to as the ZS problem, consists in finding the so called 
\emph{scattering coefficients} which are defined through special solutions 
of~\eqref{eq:ZS-prob} known as the \emph{Jost solutions} which are linearly 
independent solutions of~\eqref{eq:ZS-prob} with a prescribed behavior at 
$+\infty$ or $-\infty$. The Jost solutions of the \emph{first kind}, denoted
by $\vs{\psi}(t;\zeta)$ and $\ovl{\vs{\psi}}(t;\zeta)$, are the linearly
independent solutions of~\eqref{eq:ZS-prob} which have the following asymptotic 
behavior as $t\rightarrow\infty$:
$\vs{\psi}(t;\zeta)e^{-i\zeta t}\rightarrow(0,1)^{\tp}$ and 
$\ovl{\vs{\psi}}(t;\zeta)e^{i\zeta t}\rightarrow (1,0)^{\tp}$. The Jost solutions 
of the \emph{second kind}, denoted by
$\vs{\phi}(t;\zeta)$ and $\overline{\vs{\phi}}(t;\zeta)$, are the linearly 
independent solutions of~\eqref{eq:ZS-prob} which have the following asymptotic 
behavior as $t\rightarrow-\infty$: 
$\vs{\phi}(t;\zeta)e^{i\zeta t}\rightarrow(1,0)^{\tp}$ and 
$\overline{\vs{\phi}}(t;\zeta)e^{-i\zeta t}\rightarrow(0,-1)^{\tp}$.
The scattering coefficients are defined by 
\begin{equation}\label{eq:wrons-scoeff}
\begin{split}
&\mathfrak{a}(\zeta)= \Wrons\left(\vs{\phi},{\vs{\psi}}\right),\quad 
 \mathfrak{b}(\zeta)= \Wrons\left(\ovl{\vs{\psi}},\vs{\phi}\right),\\
&\ovl{\mathfrak{a}}(\zeta)=\Wrons\left(\ovl{\vs{\phi}},\overline{\vs{\psi}}\right),\quad
\ovl{\mathfrak{b}}(\zeta) =\Wrons\left(\ovl{\vs{\phi}},{\vs{\psi}}\right),
\end{split}
\end{equation}
for $\zeta\in\field{R}$. The analytic continuation of the Jost solution 
with respect to $\zeta$ is
possible provided the potential is decays sufficiently fast or has a compact
support. If the potential has a compact support, the Jost solutions have analytic 
continuation in the entire complex plane. Consequently, the scattering 
coefficients $\mathfrak{a}(\zeta)$, $\mathfrak{b}(\zeta)$, 
$\ovl{\mathfrak{a}}(\zeta)$, $\ovl{\mathfrak{b}}(\zeta)$ are analytic 
functions of $\zeta\in\field{C}$~\cite{AKNS1974}. 

In general, the nonlinear Fourier spectrum for the signal $q(t)$ comprises 
a \emph{discrete} and a \emph{continuous spectrum}. The discrete spectrum consists 
of the so called \emph{eigenvalues} $\zeta_k\in\field{C}_+$, such that 
$\mathfrak{a}(\zeta_k)=0$, and, the \emph{norming constants} $b_k$ such that 
$\vs{\phi}(t;\zeta_k)=b_k\vs{\psi}(t;\zeta_k)$. For compactly supported potentials, 
$b_k=b(\zeta_k)$. The continuous spectrum, also 
referred to as the \emph{reflection coefficient}, is 
defined by $\rho(\xi)={\mathfrak{b}(\xi)}/{\mathfrak{a}(\xi)}$ for $\xi\in\field{R}$.

\section{The numerical scheme}
Introducing the ``local'' scattering coefficients $a(t;\zeta)$ and 
$\breve{b}(t;\zeta)$ such that 
$\vs{\phi}(t;\zeta)e^{i\zeta t}=(a(t;\zeta), \breve{b}(t;\zeta))^{\tp}$, 
the ZS scattering problem can be written as 
$\partial_{t}a(t;\zeta)=q(t)\breve{b}(t;\zeta)$ and 
$\partial_{t}\breve{b}(t;\zeta)-2i\zeta\breve{b}(t;\zeta)=r(t)a(t;\zeta)$. Let 
the scattering potential $q(t)$ be supported in 
$\field{I}=[-1,1]$. Such signals have been studied in more detail
in~\cite{V2017INFT1,V2018TL,V2018CNSNS,V2019DT} also in order to understand the consequence 
of domain truncation for general signals. For $\zeta\in\field{C}_+$, the 
`initial' conditions for the Jost solution $\vs{\phi}$ 
are: $a(-1;\zeta) = 1$ and $\breve{b}(-1;\zeta) = 0$. The scattering coefficients 
$\mathfrak{a}$ and $\mathfrak{b}$ are given by $\mathfrak{a}(\zeta)=a(+1;\zeta)$ 
and $\mathfrak{b}(\zeta)=\breve{b}(+1;\zeta)e^{-2i\zeta}$. In the following, we describe 
a numerical scheme based on the Chebyshev 
polynomials to solve the coupled Volterra integral equations 
\begin{equation}\label{eq:Volterra-ZS}
\begin{split}
&a(t;\zeta)=1+\int_{-1}^tq(s)\breve{b}(s;\zeta)ds,\\
&\breve{b}(t;\zeta)=2i\zeta\int^t_{-1}\breve{b}(s;\zeta)ds+\int_{-1}^t r(s)a(s;\zeta)ds,
\end{split}
\end{equation} 
which are equivalent to the ZS problem with the aforementioned initial 
conditions. The numerical scheme computes the approximations to the Jost 
solutions in terms of the Chebyshev polynomials which can then used to compute 
the scattering coefficients. In the following, for the sake of brevity of presentation, 
we assume that the eigenvalues are known. The method of computing the eigenvalues 
using a Chebyshev based spectral would be presented in a future publication. Here we 
focus entirely on the computation of the norming constants. In principle, the norming 
constants can be computed simply by
evaluating the numerical approximation to $\mathfrak{b}(\zeta)$ at the eigenvalue 
$\zeta_k$ to obtain $b_k$, however, $\breve{b}(+1;\zeta_k)$ becomes negligibly small which must now be 
multiplied with $\exp(-2i\zeta_k)$ which grows exponentially. These intermediate 
quantities can easily suffer from lack of precision leading to inaccurate 
determination of $b_k$.

Let $c(t)=\sum_{n=0}^{\infty}C_nT_n(t)$. Following~\cite{V2019SNFT}, the integral 
operator $\OP{K}$ defined by $\OP{K}[c](t)=\int_{-1}^tc(s)ds=d(t)$ in the Chebyshev basis 
is given by
\begin{multline}
d(t)=\left[C_0-\frac{1}{4}C_1-\sum_{n=2}^{\infty}\frac{(-1)^{n}C_n}{n^2-1}\right]T_0(t)\\
+\left[C_0-\frac{1}{2}C_2\right]T_1(t)+
\sum_{n=2}^{\infty}\frac{1}{2n}\left[C_{n-1}-C_{n+1}\right]T_n(t).
\end{multline}
In the matrix form, $\OP{K}$ has the representation
\begin{equation}
\mathcal{K}=
\begin{pmatrix}
 1 &-\frac{1}{4} &  -\frac{1}{3} & +\frac{1}{8} & -\frac{1}{15}  & \ldots\\
 1 &          0  & -\frac{1}{2}&  &   & \\
   &   \frac{1}{4}&          0 & -\frac{1}{4} &            & \\
   &             &  \frac{1}{6} &      0   &-\frac{1}{6}   &\\
   &             &             & \ddots       &\ddots   &\ddots
\end{pmatrix}.
\end{equation}
The next step in the discretization of~\eqref{eq:Volterra-ZS} involves expanding the signal 
in the Chebyshev basis. Let 
$q(t) = \sum_{n=0}^{\infty}Q_nT_n(t)$ and $r(t) = \sum_{n=0}^{\infty}R_nT_n(t)$ 
where $R_n=-Q^*_n$. A truncated expansion upto $N$ terms can be accomplished by 
sampling the potentials at the Chebyshev--Gauss--Lobatto (CGL) nodes given by 
$t_n = -\cos[n\pi/(N-1)],\,n=0,1,\ldots N-1$ and carrying out discrete Chebyshev transform 
which can be implemented using an FFT of size $2(N-1)$~\cite{G2011}. 

\begin{figure}[!ht]
\begin{center}
\includegraphics[scale=1]{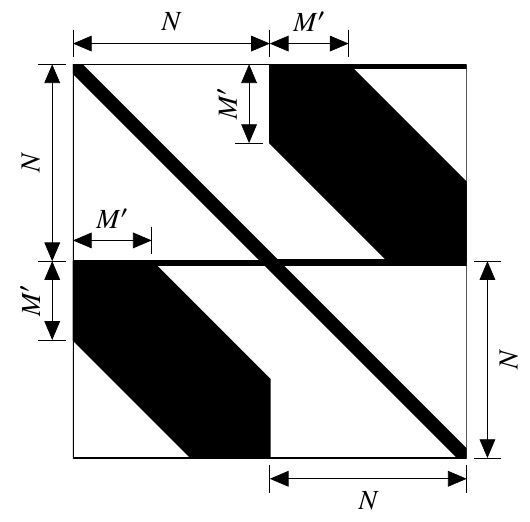}%
\caption{\label{fig:spy-mat}The sparsity pattern of a truncated 
version of the matrix $\mathcal{S}$ where $M'=M+1$ with $M$ being the number 
of Chebyshev polynomials used for approximating $q(t)$.}
\end{center}
\end{figure}

Now, our final goal is to 
obtain an expansion of the local scattering coefficients in the Chebyshev basis: To this end, let
$a(t;\zeta)= \sum_{n=0}^{\infty}A_nT_n(t)$ and $\breve{b}(t;\zeta)= 
\sum_{n=0}^{\infty}B_nT_n(t)$ where $A_n$ and 
$B_n$ are to be determined (for fixed value of $\zeta$). The last ingredient 
needed in the discretization of~\eqref{eq:Volterra-ZS} are the products $r(t)a(t;\zeta)$ and 
$q(t)\breve{b}(t;\zeta)$ which must be represented as linear operations on the unknown coefficient vectors
$\vv{A}=(A_0,A_1,\ldots)^{\tp}$ and $\vv{B}=(B_0,B_1,\ldots)^{\tp}$. Again 
following~\cite{V2019SNFT}, let
$r(t)a(t;\zeta)= \sum_{l=0}^{\infty}G_lT_l(t)$ and $q(t)\breve{b}(t;\zeta)= \sum_{l=0}^{\infty}H_lT_l(t)$;
then, $2G_0 = 2R_0A_0 +\sum_{k=1}^{\infty}R_kA_k$,
$2H_0 = 2Q_0B_0 +\sum_{k=1}^{\infty}Q_kB_k$ and
\begin{equation}
\begin{split}
2G_l &=\sum_{k=0}^{l-1}R_{l-k}A_k
+2R_{0}A_{l}+\sum_{k=1}^{\infty}R_{k}A_{k+l}
+\sum_{k=0}^{\infty}R_{k+l}A_k,\\
2H_l &=\sum_{k=0}^{l-1}Q_{l-k}B_k
+2Q_0B_l+\sum_{k=1}^{\infty}Q_{k}B_{l+k}
+\sum_{k=0}^{\infty}Q_{k+l}B_k.
\end{split}
\end{equation}
for $l\in\field{N}$. These relations 
define the operator $\mathcal{M}[\vv{Q}]$, which comprises a T\"oplitz and an 
almost Hankel matrix given by
\begin{equation}
2\mathcal{M}[\vv{Q}]=
\begin{pmatrix}
 2Q_0  & Q_1 &  Q_2 &\cdots\\
  Q_1 & 2Q_0 &  Q_1 &\ddots\\
  Q_2 &  Q_1 & 2Q_0 &\ddots\\
\vdots&\ddots&\ddots&\ddots
\end{pmatrix}\\
+
\begin{pmatrix}
   0  &  0   &   0  &\cdots\\
  Q_1 &  Q_2 & Q_3  &\sdots\\
  Q_2 &  Q_3 & Q_4  &\sdots\\
\vdots&\sdots&\sdots&\sdots
\end{pmatrix},
\end{equation}
Similarly, the representation of the operator $\mathcal{M}[\vv{R}]$ can be defined. Setting 
$\Lambda=\mathcal{K}\mathcal{M}[\vv{Q}]$, the discrete version 
of~\eqref{eq:Volterra-ZS} can be stated as
\begin{equation}\label{eq:Volterra-ZS-discrete}
\begin{pmatrix}
I & -\Lambda\\
\Lambda^* & I-2i\zeta\mathcal{K}
\end{pmatrix}
\begin{pmatrix}
\vv{A}\\
\vv{B}
\end{pmatrix}=
\mathcal{S}
\begin{pmatrix}
\vv{A}\\
\vv{B}
\end{pmatrix}=
\begin{pmatrix}
\vv{E}_0\\
\vv{0}
\end{pmatrix},
\end{equation}
where $I$ is the identity matrix, $\vv{E}_0=(1,0,\ldots)^{\tp}$ and 
$\vv{0}=(0,0,\ldots)^{\tp}$. Noting that $\vv{A}=\vv{E}_0+\Lambda\vv{B}$ and 
setting $\Gamma =\Lambda^*\Lambda$, we 
have $\left(I-2i\zeta\mathcal{K}+\Gamma\right)\vv{B}=\mathcal{K}\vv{R}$ where we 
have used the fact that $\mathcal{M}[\vv{R}]\vv{E}_0=\vv{R}$. The numerical scheme 
can be obtained as follows: truncation of the Chebyshev 
expansion of $q(t)$ to $M$ terms, truncation of $\vv{E}_0$, $\vv{A}$ and $\vv{B}$ to $N$-dimensional 
vectors, and, truncation of $\Lambda$ and $\mathcal{K}$ to $N\times N$ matrix where 
$N\geq2M$. If a direct sparse solver is used, the complexity of the numerical scheme 
would be lower than $\bigO{N^3}$. 
\begin{rem}
Let us remark that within the iterative approach and using the structured nature of the
matrices involved it is possible to lower the complexity of the linear solver as 
in~\cite{V2019SNFT}. Our preliminary investigation indicate that the formulation 
\begin{equation}
\left[I+\left(I-2i\zeta\mathcal{K}\right)^{-1}\Gamma\right]\vv{B}
=\left(I-2i\zeta\mathcal{K}\right)^{-1}\mathcal{K}\vv{R},
\end{equation}
works better for iterative solvers. However, the number of iterations needed for 
the stabilized biconjugate gradient method~\cite{V1992} can be as large as $50$. Let us also 
mention that fast versions of the direct solvers for banded systems with certain number of 
filled first rows have been proposed~\cite{OT2013}. We defer these aspects to a future publication.
\end{rem}

The recipe discussed above accomplishes 
the computation of the Jost function of second kind $\vs{\phi}$, let us now show that by solving 
the scattering problem in the setting described above for $q^*(-t)$ it is possible 
to compute the Jost solution of the first kind $\vs{\psi}$. If $\vs{\Phi}(t;\zeta)$ 
denotes the Jost solution of the second kind for $q^*(-t)$, we have 
$\vs{\psi}(t;\zeta)=(\Phi_2(-t;\zeta), \Phi_1(-t;\zeta))^{\tp}$~\cite{V2017INFT1}. For convenience, 
let $\vs{\psi}(t;\zeta)e^{-i\zeta t}=(\breve{c}(t;\zeta),d(t;\zeta))^{\tp}$ so that, 
for $\zeta\in\field{C}_+$, the `initial' conditions are: $d(1;\zeta) = 1$ and 
$\breve{c}(1;\zeta) = 0$. The scattering coefficients 
$\mathfrak{a}$ and $\ovl{\mathfrak{b}}$ are given by $\mathfrak{a}(\zeta)=d(-1;\zeta)$ 
and $\ovl{\mathfrak{b}}(\zeta)=\breve{c}(-1;\zeta)e^{-2i\zeta}$. Again, we do not attempt 
to evaluate $\ovl{\mathfrak{b}}$ at the eigenvalues $\zeta_k$.  Finally, let us note that the Chebyshev 
coefficients of $q^*(-t)$ are given by $(-1)^nQ_n,\,n\in\field{N}_0,$ which follows 
from the symmetry property of the Chebyshev polynomials: $T_n(-t)=(-1)^nT_n(t)$. 


\begin{figure}[!h]
\centering
\includegraphics[scale=1]{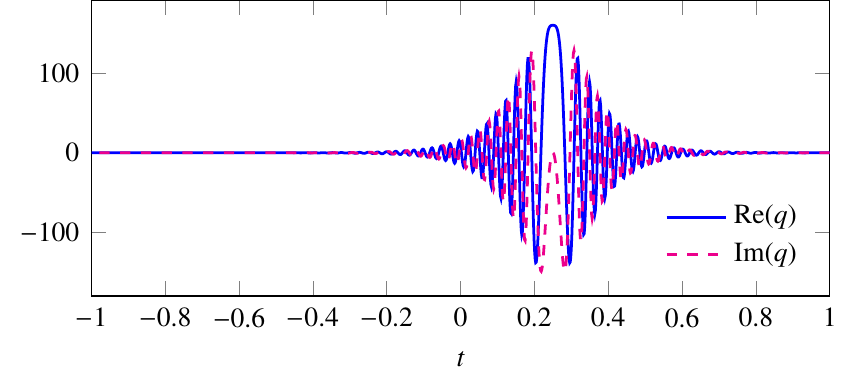}%
\caption{\label{fig:sig0} The figure shows a chirped hyperbolic signal for which 
the parameters are $\mu=0.8$, $K=8$, $A_0=K/\lambda\approx16.66$ and 
$t_0=0.25$. The scaling parameter is $W=12$.}
\end{figure}
\begin{figure}[!h]
\centering
\includegraphics[scale=1]{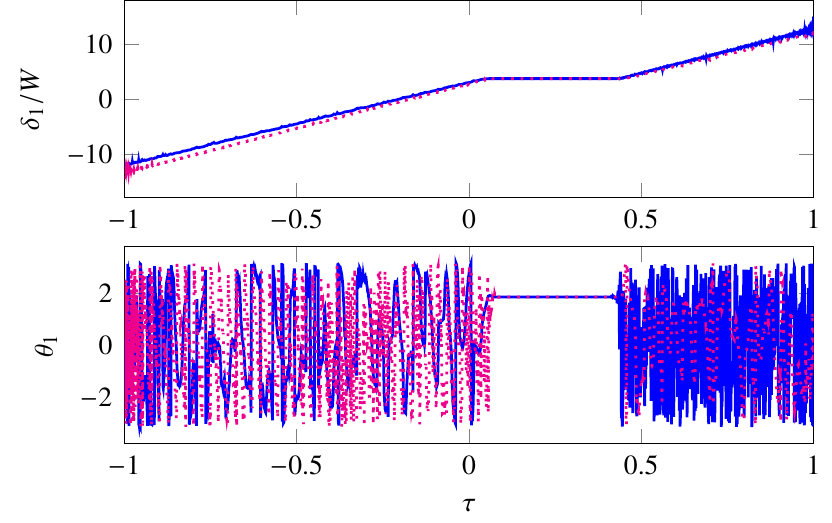}%
\caption{\label{fig:nconst0} The figure shows $\delta_1$ (top) and 
$\theta_1$ (bottom) computed for the signal in Fig.~\ref{fig:sig0} from the expressions defined 
in~\eqref{eq:theta-delta}, namely, $f_{\delta}$ (solid line) and $g_{\delta}$ (dotted line) 
for $\delta_1$, and, $f_{\theta}$ (solid line), $g_{\theta}$ (dotted line) for $\theta_1$ where 
the corresponding eigenvalue is $\zeta_1=i9.5W$. The MTV algorithm (applied to 
$f_{\delta}$ and $g_{\delta}$) with sliding-window size 
$m=20$ yields $\tau=0.24,0.26$ for which the errors are as follows: 
$|\delta_1-\delta^{(\text{num.})}_1|=1.16\times 10^{-7},1.16\times10^{-7}$, 
$|\theta_1-\theta_1^{(\text{num.})}|=1.77\times10^{-7},1.77\times10^{-7}$, respectively.}
\end{figure}
\begin{figure*}[!th]
\centering
\includegraphics[scale=1]{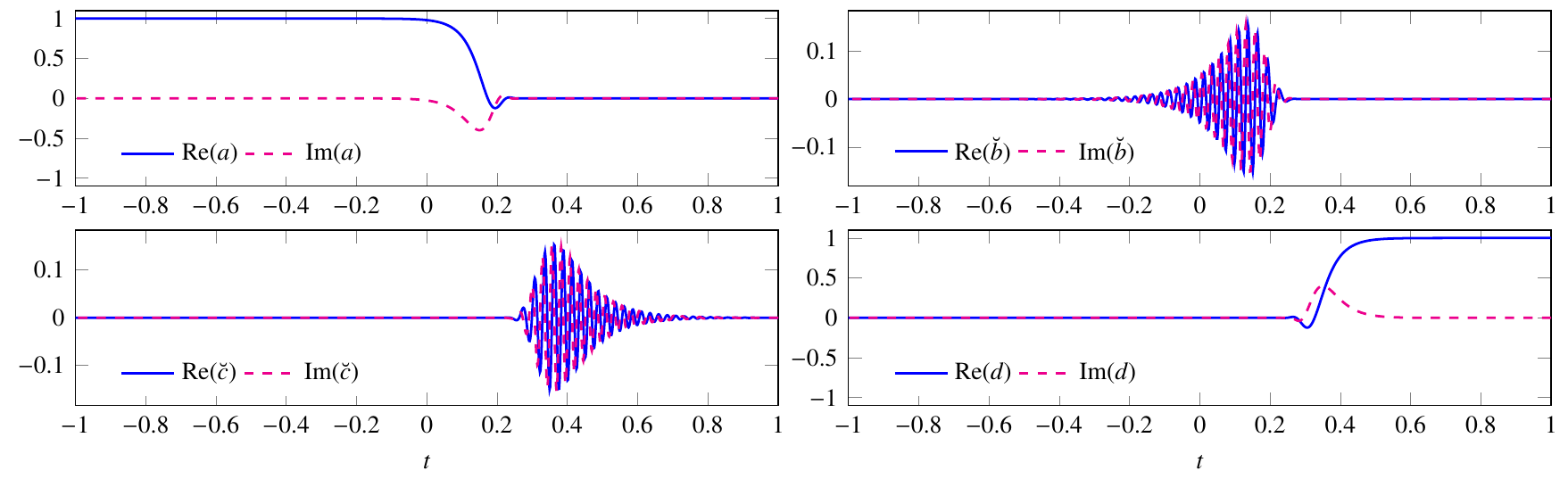}%
\caption{\label{fig:jost0} The figure shows the numerically computed (modified) 
Jost solutions for the signal shown in Fig.~\ref{fig:sig0} corresponding to 
the eigenvalue $\zeta_1=i9.5W$ with numerical parameters given by $M=2^9$ 
and $N=2^{11}$. The quality of the approximation can be assessed from 
$|\mathfrak{a}^{(\text{num.})}(\zeta_1)|\approx3.36\times10^{-16}$.}
\end{figure*}

Having discussed the computation of the Jost solutions, we turn to the computation 
of the norming. To this end, let $\zeta_k=i\xi_k+\eta_k$. For convenience, we set 
$b_k=\exp(i\theta_k+\delta_k)$ and introduce the functions
\begin{equation}\label{eq:theta-delta-func}
\begin{split}
&f_{\delta}(\tau;\zeta_k)
=\ln\left|\frac{a(\tau;\zeta_k)}{\breve{c}(\tau;\zeta_k)}\right|+2\Im{\zeta_k}\tau,\\ 
&g_{\delta}(\tau;\zeta_k)
=\ln\left|\frac{\breve{b}(\tau;\zeta_k)}{d(\tau;\zeta_k)}\right|+2\Im{\zeta_k}\tau,\\
&f_{\theta}(\tau;\zeta_k)
=\arg\left[\frac{a(\tau;\zeta_k)}{\breve{c}(\tau;\zeta_k)}e^{-2i\Re{\zeta_k}\tau}\right]\\
&g_{\theta}(\tau;\zeta_k)
=\arg\left[\frac{\breve{b}(\tau;\zeta_k)}{d(\tau;\zeta_k)}e^{-2i\Re{\zeta_k}\tau}\right],
\end{split}
\end{equation}
so that
\begin{equation}\label{eq:theta-delta}
\delta_k=f_{\delta}(\tau;\zeta_k)=g_{\delta}(\tau;\zeta_k),\quad
\theta_k=f_{\theta}(\tau;\zeta_k)=g_{\theta}(\tau;\zeta_k),
\end{equation}
for $\tau\in\field{I}$. The functions defined above are constant with respect to $\tau$ 
for any given eigenvalue $\zeta_k$, however, at the discrete level they may vary. The choice 
of $\tau$ where the expressions above must be evaluated depends on the numerical conditioning 
of the quantities involved. Let a sub-grid $J$ be said to be admissible with respect to 
$f_{\delta}$ if its \emph{total variation} over $J$, given by 
\begin{equation}
\begin{split}
\OP{V}[f_{\delta};J]=
\sum_{n}|f_{\delta}(\tau_{n+1};\zeta_k)-f_{\delta}(\tau_{n};\zeta_k)|,
\end{split}
\end{equation}
satisfies the condition $\OP{V}[f_{\delta};J]\leq\epsilon, (\epsilon>0)$. In the numerical implementation, 
we introduce a sliding window given by $J_n(m)=\{\tau_n,\tau_{n+1},\ldots,\tau_{n+m-1}\}$ 
where $m<N$ is fixed. If an appropriate tolerance $\epsilon$ cannot be guessed 
\textit{a priori}, we simply choose $J_m(n)$ that corresponds to 
$\min_n\OP{V}[f_{\delta};J_m(n)]$. The point $\tau_{n+m/2}$ can then be reported as 
the optimal point for the computation of $\delta$ and $\theta$. We label this algorithm as the 
\emph{minimum} total variation (MTV) algorithm. 

\section{Numerical Tests}
For numerical tests, we consider the chirped secant-hyperbolic potential~\cite{TVZ2004} given by
$q(t)=Wf((t-t_0)/W)$ for $t\in\field{I}$ where 
\begin{equation}
f(s) = A_0{\exp[-2i\mu A_0\log(\cosh s)]}/{\cosh(s)},\quad s\in\field{R}.
\end{equation} 
Here, $W>0$ controls how well the signal $q(t)$ is supported in $\field{I}$. Let 
$\mu\in[0,1)$ and $\lambda=\sqrt{1-\mu^2}\in(0,1]$ and set 
$\omega=\lambda+i\mu\in\field{T}$. Further, set $\tilde{A}_0=\lambda A_0+{1}/{2}$ and let 
$K=\floor{\tilde{A}_0}$; then, the eigenvalues are given by 
$\zeta_k = i\left(\tilde{A}_0-k\right)W$ and the corresponding norming constants are given by 
\begin{equation}
b_k = \omega e^{-2i\mu A_0(\log2)+i\pi k}
\prod_{j=1}^{k-1}\left(\frac{\omega A_0-j}{\omega^*A_0-j}\right)e^{2i\zeta_kt_0},
\end{equation}
for $k=1,2,\ldots,K$. We set $\tilde{A}_0=K+1/2$ so that $A_0=K/\lambda$ and 
\begin{equation}
\zeta_k = i\left(K+1/2-k\right)W,\quad k=1,2,\ldots,K.
\end{equation}
The signal for the choice of parameters $W=12$, $\mu=0.8$, $t_0=0.25$ and $K=8$ 
is shown in Fig.~\ref{fig:sig0}. The two expression for $\delta_1$ and $\theta_1$ 
provided in~\eqref{eq:theta-delta} are plotted in Fig.~\ref{fig:nconst0} which 
correspond to the numerically computed Jost solution for $\zeta_1$ shown in 
Fig.~\ref{fig:jost0} with $M=2^9$ and $N=4M$. It is evident from Fig.~\ref{fig:nconst0} 
that the choice of $\tau$ is non-trivial and the choice of $\tau=0$ is certainly not 
admissible. The MTV algorithm applied to the functions 
$f_{\delta}$ and $g_{\delta}$ with sliding-window size $m=20$ finds the optimal points to be 
$\tau=0.24,0.26$, respectively. The errors $|\delta_1-\delta^{(\text{num.})}_1|$ and 
$|\theta_1-\theta_1^{(\text{num.})}|$ for each of the choices of $\tau$ are of the order 
$10^{-7}$.

\begin{figure}[!h]
\centering
\includegraphics[scale=1]{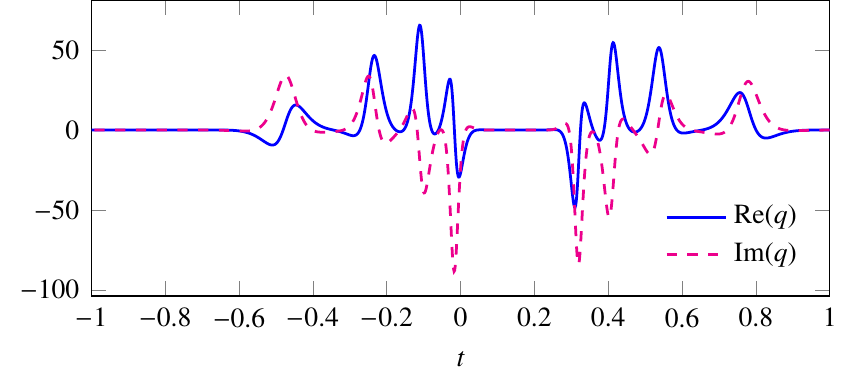}%
\caption{\label{fig:sig1} The figure shows the $8$-soliton solution 
corresponding to the discrete spectrum listed in 
Table~\ref{tb:8-soliton}. The scaling parameter is $W=10$.}
\end{figure}

\begin{figure*}[!th]
\centering
\includegraphics[scale=1]{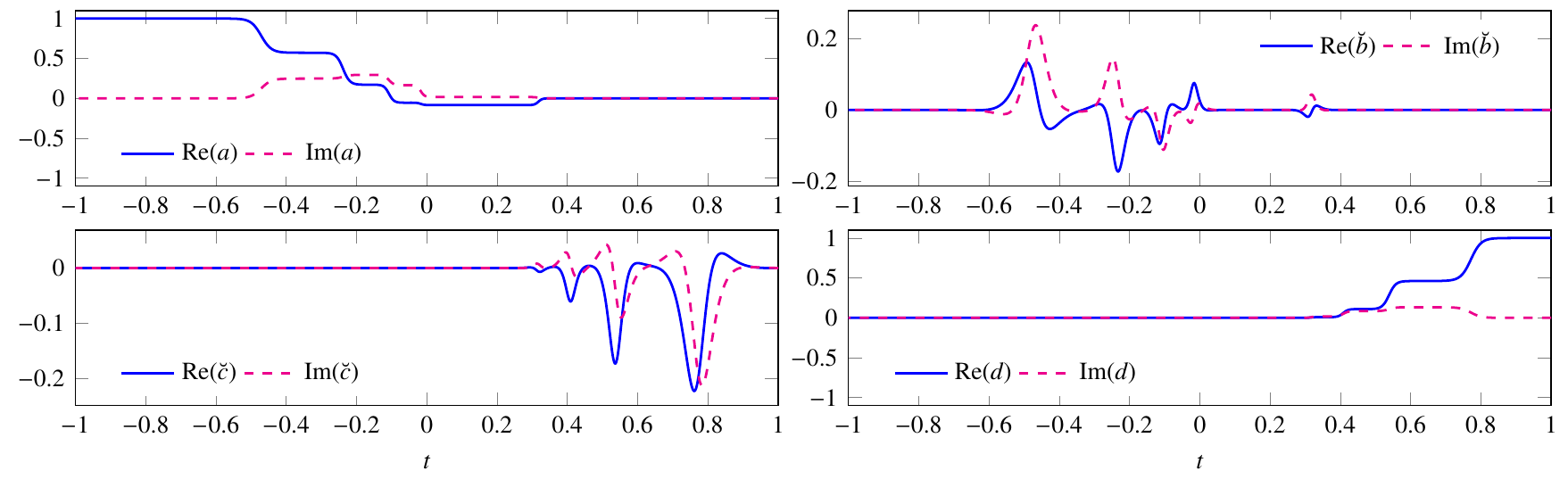}%
\caption{\label{fig:jost1} The figure shows the numerically computed (modified) 
Jost solutions for the signal shown in Fig.~\ref{fig:sig1} corresponding to 
the eigenvalue $\zeta_1/W=-2.5 +4.33013i$ with numerical parameters given by 
$M=2^9$ and $N=2^{11}$. The quality of the approximation can be assessed from 
$|\mathfrak{a}^{(\text{num.})}(\zeta_1)|\approx2.99\times10^{-10}$.}
\end{figure*}

\begin{figure}[!h]
\centering
\includegraphics[scale=1]{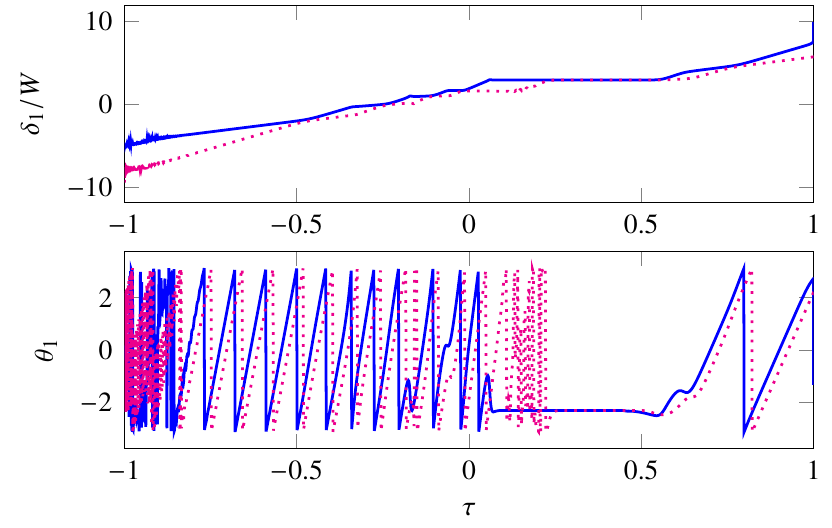}%
\caption{\label{fig:nconst1} The figure shows $\delta_1$ (top) and 
$\theta_1$ (bottom) computed for the signal in Fig.~\ref{fig:sig1} from the expressions defined 
in~\eqref{eq:theta-delta}, namely, $f_{\delta}$ (solid line) and $g_{\delta}$ (dotted line) 
for $\delta_1$, and, $f_{\theta}$ (solid line), $g_{\theta}$ (dotted line) for $\theta_1$ where 
the corresponding eigenvalue is $\zeta_1/W=-2.5 +4.33013i$. The MTV algorithm (applied to 
$f_{\delta}$ and $g_{\delta}$) with sliding-window size 
$m=20$ yields $\tau\approx0.29,0.34$ for which the errors are as follows: 
$|\delta_1-\delta^{(\text{num.})}_1|=7.74\times 10^{-7},5.29\times10^{-7}$,
$|\theta_1-\theta_1^{(\text{num.})}|=4.74\times10^{-7},8.51\times10^{-7}$, 
respectively.}
\end{figure}

A second example to demonstrate the effectiveness of the MTV algorithm is that of an 
$8$-soliton solution whose discrete spectrum in terms of the triplets 
$(\zeta_k,\delta_k,\theta_k)$ is listed in Table~\ref{tb:8-soliton}. The signal computed 
using the classical Darboux transformation~\cite{VW2016OFC,V2017INFT1,V2018CNSNS,V2019DT} for the 
choice $W=10$ is shown in Fig.~\ref{fig:sig1} and numerically computed Jost solutions 
in Fig.~\ref{fig:jost1}. The two expression for $\delta_1$ and $\theta_1$ 
provided in~\eqref{eq:theta-delta} are plotted in Fig.~\ref{fig:nconst1} which 
correspond to the numerically computed Jost solution for $\zeta_1$ 
with $M=2^9$ and $N=4M$. It is evident from Fig.~\ref{fig:nconst1} that the 
choice of $\tau$ is again non-trivial and the choice of $\tau=0$ is once again not 
admissible. The MTV algorithm applied to the functions 
$f_{\delta}$ and $g_{\delta}$ with sliding-window size $m=20$ finds the optimal points to be 
$\tau\approx0.29,0.34$, respectively. The errors $|\delta_1-\delta^{(\text{num.})}_1|$ and 
$|\theta_1-\theta_1^{(\text{num.})}|$ for each of the choices of $\tau$ are of the order 
$10^{-7}$.

\begin{figure}[!h]
\centering
\includegraphics[scale=1]{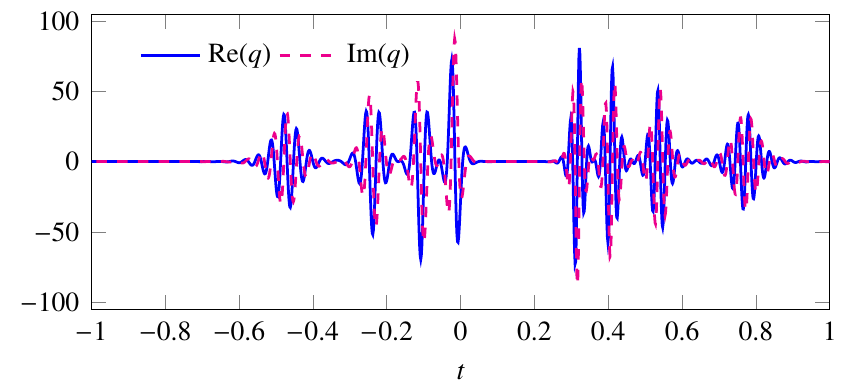}%
\caption{\label{fig:sig2} The figure shows a signal derived from the 
$8$-soliton solution corresponding to the discrete spectrum listed in 
Table~\ref{tb:8-soliton} by multiplying a linear phase factor given by 
$\exp(i8\Re(\zeta_1)t)$. The scaling parameter is $W=10$.}
\end{figure}

\begin{figure}[!h]
\centering
\includegraphics[scale=1]{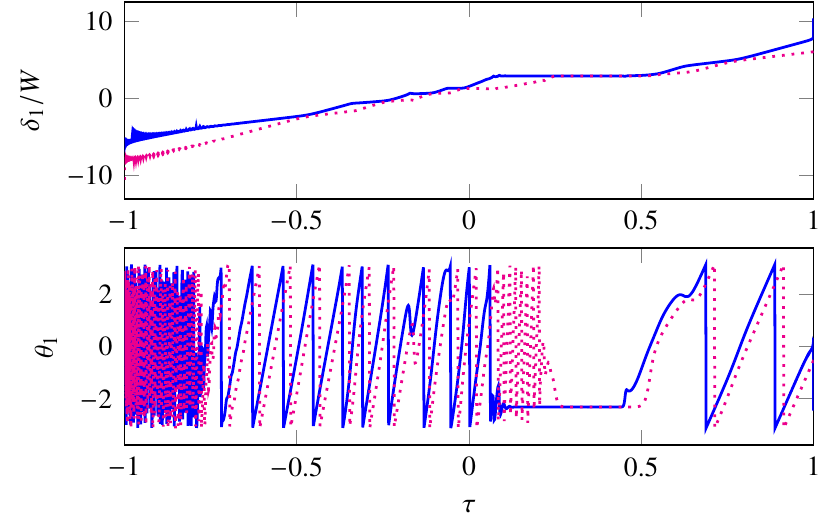}%
\caption{\label{fig:nconst2} The figure shows $\delta_1$ (top) and 
$\theta_1$ (bottom) computed for the signal in Fig.~\ref{fig:sig2} from the expressions defined 
in~\eqref{eq:theta-delta}, namely, $f_{\delta}$ (solid line) and $g_{\delta}$ (dotted line) 
for $\delta_1$, and, $f_{\theta}$ (solid line), $g_{\theta}$ (dotted line) for $\theta_1$ where 
the corresponding eigenvalue is $\zeta_1/W=8.5 +4.33013i$. The MTV algorithm (applied to 
$f_{\delta}$ and $g_{\delta}$) with sliding-window size 
$m=20$ yields $\tau\approx0.25,0.36$ for which the errors are as follows: 
$|\delta_1-\delta^{(\text{num.})}_1|=3.11\times 10^{-4},2.96\times10^{-4}$,
$|\theta_1-\theta_1^{(\text{num.})}|=9.64\times10^{-7},2.44\times10^{-5}$, 
respectively.}
\end{figure}

\begin{figure*}[!th]
\centering
\includegraphics[scale=1]{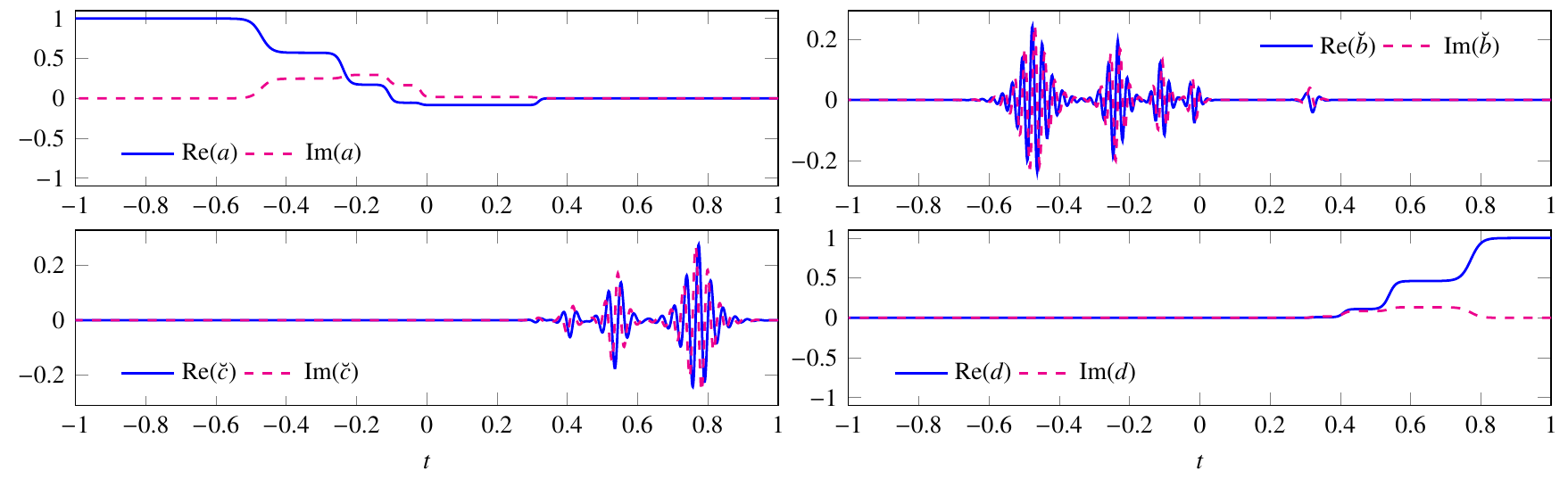}%
\caption{\label{fig:jost2} The figure shows the numerically computed (modified) 
Jost solutions for the signal shown in Fig.~\ref{fig:sig2} corresponding to 
the eigenvalue $\zeta_1/W=8.5 +4.33013i$. The quality of the approximation can be 
assessed from $|\mathfrak{a}^{(\text{num.})}(\zeta_1)|\approx8.20\times10^{-9}$.}
\end{figure*}

The third example to demonstrate the effectiveness of the MTV algorithm is derived from 
the last example by multiplying a linear phase factor $\exp(i2\xi_{\text{shift}}t)$ to the signal 
so that the eigenvalues acquire a shift of $\xi_{\text{shift}}$ which we choose 
to set $4\Re{\zeta_1}$ (see Table~\ref{tb:8-soliton}). We let the numerical parameters 
to be the same as in the last example. The signal shown in Fig.~\ref{fig:sig2}, 
the variation of $\delta_1,\theta_1$ are shown in Fig.~\ref{fig:nconst2} which 
correspond to the numerically computed Jost solution (see Fig.~\ref{fig:jost2}) 
for $\zeta_1+4\Re{\zeta_1}$. The 
MTV algorithm applied to the functions $f_{\delta}$ and $g_{\delta}$ finds the optimal 
points to be $\tau\approx0.25,0.36$, respectively. The errors 
$|\delta_1-\delta^{(\text{num.})}_1|$ and $|\theta_1-\theta_1^{(\text{num.})}|$ for 
each of the choices of $\tau$ are of the order $10^{-4}$ and $10^{-5}$, respectively.

\begin{table}[tbph!]
\def\arraystretch{1.75}
\caption{Discrete spectrum\label{tb:8-soliton}}
\begin{center}
\begin{tabular}{c|c|c|c}
$k$ & $\zeta_k/W$ & $\theta_k$ & $\delta_k$\\
\hline
$1$ & $-2.5 +4.33013i$ & $2.74889 $ & $+16$\\
$2$ & $+2.5 +4.33013i$ & $2.35619 $ & $-16$\\
$3$ & $-2   +3.4641i $ & $1.9635  $ & $+16$\\
$4$ & $+2   +3.4641i $ & $1.5708  $ & $-16$\\
$5$ & $-1.5 +2.59808i$ & $1.1781  $ & $+16$\\
$6$ & $+1.5 +2.59808i$ & $0.785398$ & $-16$\\
$7$ & $-1   +1.73205i$ & $0.392699$ & $+16$\\
$8$ & $+1   +1.73205i$ & $0       $ & $-16$\\
\hline
\end{tabular}
\end{center}
\end{table}


The final numerical tests are meant to verify the spectral convergence 
of the numerical scheme. To this end, we quantify the error by
\begin{equation}\label{eq:error-metric}
e_{\theta}=\frac{1}{K}\sum_{k=1}^K|\theta_k-\theta^{\text{num.}}_k|,\quad 
e_{\delta}=\frac{1}{K}\sum_{k=1}^K|\delta_k-\delta^{\text{num.}}_k|.
\end{equation}
We set $N=4M$ and consider the set of values $K\in\{4,8,12,16\}$. 

\begin{figure}[!th]
\centering
\includegraphics[scale=1]{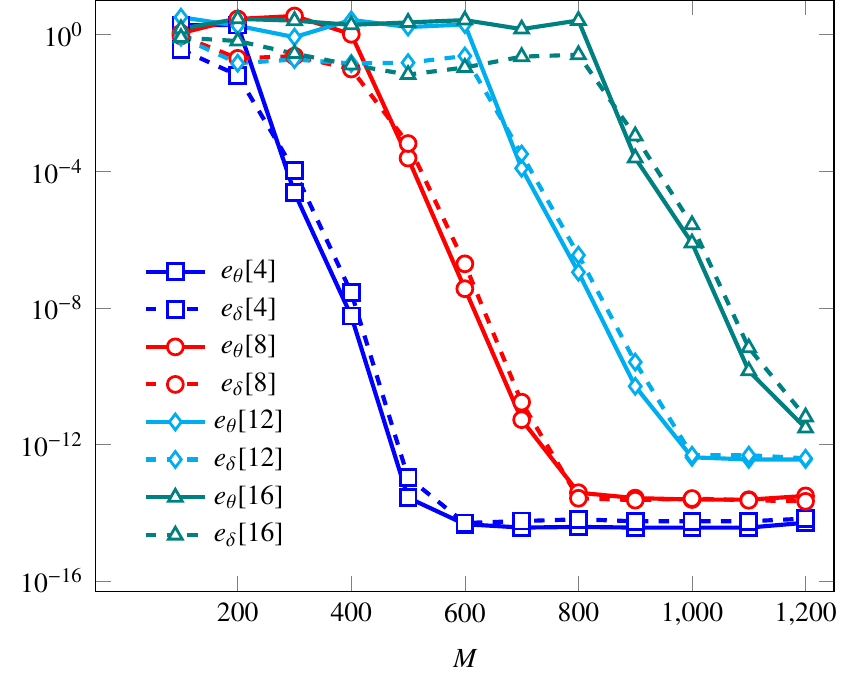}%
\caption{\label{fig:nconst-convg}The figure shows the convergence analysis 
for the chirped secant hyperbolic profile with $\mu=0.8$, $t_0=0$ and the 
scaling parameter $W=20$. The legends depict the quantities $e_{\delta}[K]$ 
and $e_{\theta}[K]$ defined by~\eqref{eq:error-metric} and parametrized by 
$K$, the number of eigenvalues.}
\end{figure}

The first profile we would like to use for the 
convergence analysis is chirped hyperbolic profile. Let us set the parameters as 
$\mu=0.8$, $t_0=0$ and $W=20$. The choice of these 
parameters makes $\tau=0$ optimal for the computation of the norming 
constants. The results of the convergence analysis is shown in 
Fig.~\ref{fig:nconst-convg} which confirms the spectral convergence 
of the numerical scheme. The plateauing of the error seen in these 
plots are on account of the lack of compact support of $q(t)$. 

\begin{figure}[!ht]
\begin{center}
\includegraphics[scale=1]{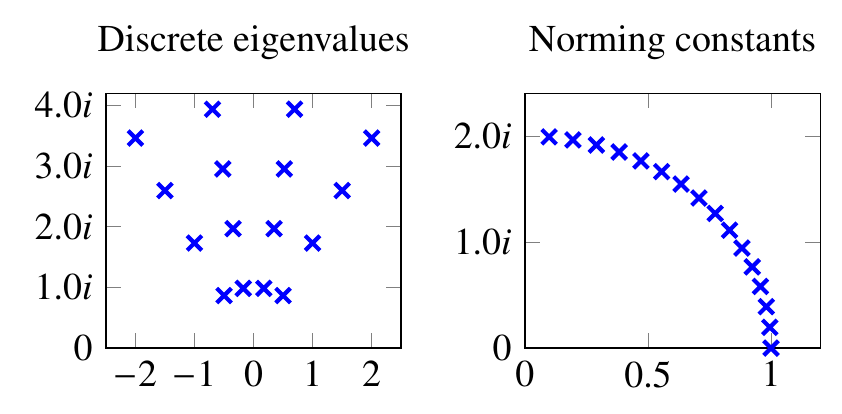}%
\caption{\label{fig:MS-spectrum}The figure shows the discrete spectrum of a multisoliton solution.}
\end{center}
\end{figure}

\begin{figure}[!th]
\centering
\includegraphics[scale=1]{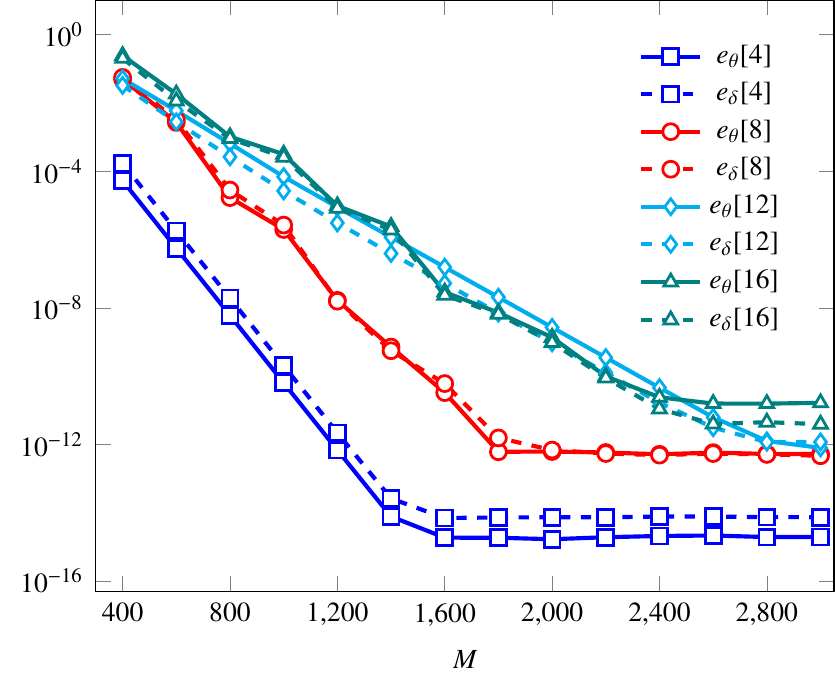}%
\caption{\label{fig:nconst-nsoliton}The figure shows the convergence analysis 
for multisoliton solutions with number of eigenvalues $K=4,8,12,16$ where the discrete 
spectrum in each of the cases is a subset of that shown in Fig.~\ref{fig:MS-spectrum}. The legends 
depict the quantities $e_{\delta}[K]$ and $e_{\theta}[K]$ defined 
by~\eqref{eq:error-metric} and parametrized by $K$.}
\end{figure}

The second profile we would like to use for the 
convergence analysis are multisoliton solutions. Let $\theta_j=\pi(j+2)/9\in[\pi/3,2\pi/3]$ for 
$j=1,\ldots,4$. Then the eigenvalues are defined as 
$\zeta_{j+4(l-1)}=l\exp(i\theta_j),\,l=1,\ldots,4,\,j=1,\ldots,4$. The norming 
constants are chosen as $b_j=\exp[i\pi (j-1)/16],\,j=1,2,\ldots,16$. The discrete spectrum 
corresponding to $K=16$ is depicted in Fig.~\ref{fig:MS-spectrum}. The potential 
can be computed with machine precision using the classical Darboux transformation 
algorithm~\cite{V2017INFT1}. The scaling parameter is set to be $W=22$. Here, 
$\tau=0$ is known to be optimal for the computation of the norming 
constants. The results of the convergence analysis is shown in 
Fig.~\ref{fig:nconst-nsoliton} which confirms the spectral convergence 
of the numerical scheme. The plateauing of the error seen in these plots are 
again on account of the lack of compact support of $q(t)$. 



\section{Conclusion}
In this paper, we presented a Chebyshev spectral method for the solution of the 
Zakharov--Shabat scattering problem for complex values of the spectral parameter. Within this 
discrete framework, we also proposed a robust algorithm for computing the norming constants. This 
algorithm is based on a minimum total variation principle and therefore the algorithm is 
abbreviated as the MTV algorithm for norming constants. Future work in this direction will 
focus on developing fast solvers for the linear system within the direct (relying on the 
structured nature of the system matrix) as well as iterative (relying on the fast 
matrix--vector multiplication for structured matrices) methods with and without preconditioning.

\IEEEtriggeratref{13}

\bibliographystyle{IEEEtran}

\providecommand{\noopsort}[1]{}\providecommand{\singleletter}[1]{#1}%

\end{document}